\documentclass[reqno]{amsart}

\usepackage{amsmath, amssymb}
\usepackage{graphicx}
\usepackage{dcolumn}
\usepackage{bm}
\usepackage{hyperref}
\usepackage{setspace}
\usepackage{color}

\newcommand{\lra}[1]{\ensuremath{{\langle #1\rangle}}}

\begin{document}

\title{%
  Constraint-Preserving Scheme for Maxwell's Equations
}%

\author[T. Tsuchiya]{Takuya Tsuchiya}
\author[G. Yoneda]{Gen Yoneda}
\address[T. Tsuchiya, G. Yoneda]{
  Department of Mathematics, School of Fundamental Science and Engineering,
  Waseda University, Okubo, Shinjuku, Tokyo 169-8555, Japan}
\email[T. Tsuchiya]{t-tsuchiya@aoni.waseda.jp, tsuchiya@akane.waseda.jp}

\maketitle

\begin{abstract}
  We derive the discretized Maxwell's equations using the discrete variational
  derivative method (DVDM), calculate the evolution equation of the constraint,
  and confirm that the equation is satisfied at the discrete level.
  Numerical simulations showed that the results obtained by the DVDM are
  superior to those obtained by the Crank-Nicolson scheme.
  In addition, we study the two types of the discretized Maxwell's equations by
  the DVDM and conclude that if the evolution equation of the constraint is
  not conserved at the discrete level, then the numerical results are also
  unstable.
\end{abstract}

\section{Introduction}
Constrained dynamical systems which are expressed by a set of evolution
equations with constraints are important role in various branches of physics.
For instance, the gauge theory is one of such systems, it is well known that
the theory has been studied widely in the modern physics and mathematics.
In addition, the canonical quantization of the constrained dynamical systems
which is quantizing the Hamiltonian formalism in the classical theory was
studied mainly by Dirac \cite{Dirac}.
So the constrained dynamical systems are often formulated in the Hamiltonian
formalism.
If we analyze the constrained dynamical systems using simulations, it is not
easy to conserve the constraints in the evolution.
This is because the violations of the constraints caused by the numerical errors
are grown in long-term free evolution.
So we should make discretization of the equations using appropriate numerical
schemes.

Recently, a number of investigations of schemes to derive suitable discretized
equations have been reported.
Some of these schemes are referred to as geometrical integration \cite{HL}, in
which the numerical errors in the initial state are preserved.
The discrete variational derivative method (DVDM) is one such method.
The DVDM was proposed then extended by Mori, Furihata, Matsuo, and Yaguchi
\cite{FM96, Furihata99, MF01, FM, AMM15, IY15}.
The method can apply the set of evolution equations which is derived by the
variational principle in continuous level.
If the set of the equations in continuous level have the dissipation and
conservation properties, the set of discretized evolution equations using this
method expects to preserve these properties.
The DVDM is one of the methods of constructions of suitable numerical schemes,
it would be possible to construct constraint preserving discrtized equations
without the method.
However, the idea is similar to the variational principle which is one of the
most useful and powerful analytical tools in mathematics and physics.
Therefore, this method seems to apply the various equations which have the
Lagrangian and Hamiltonian density.

In this article, the purpose is finding the conditions to construct the
appropriate discretization methods of the constrained systems.
Especially, we target the types that the Hamiltonian density has the gauge
freedoms, in other words the Hamiltonian density includes the constraints.
So we select the Maxwell's equations as the example.
We seem the Maxwell's equations are one of appropriate examples of the
constrained dynamical systems.
There are two reasons.
First is that the equations are the linear with regard to the dynamical
variables, we study easily relatively to the nonlinear cases.
Second is that the structure is easily relatively to the one in modern physics
such as the general relativity.

We organize our article as follows.
In section \ref{sec:Maxwell}, we review the Maxwell's equations of the
Hamiltonian formalism and derive the evolution equation of the constraint.
In section \ref{sec:DVDM}, the brief reviews of the DVDM of the Hamiltonian
formalism are shown.
In section \ref{sec:discreteMaxwellSystems}, we derive the three types of the
discrete Maxwell's equations.
The one is using the ICNS, the others are using the DVDM.
In section \ref{sec:Numericaltests}, we preform some simulations using the
discretized equations made in Section \ref{sec:discreteMaxwellSystems}, we
summarize this article in Section \ref{sec:summary}.
In this article, indices such as $(i,j,k,\cdots)$ run from 1 to 3.
We use the Einstein convention of summation of repeated up-down indices.
The indices are raised and lowered by the Kronecker delta.

\section{Maxwell's Equations
  \label{sec:Maxwell}
}

We begin by introducing the Maxwell's equations and their Hamiltonian density.
The equations are (e.g., see \cite{Jackson})
\begin{align}
  \epsilon_0\partial_t E_i - \frac{1}{\mu_0}\varepsilon_i{}^{mn}\partial_m B_n
  &= -J_i,\label{eq:AM}\\
  \partial_t B_i + \varepsilon_i{}^{mn}\partial_m E_n &= 0,
  \label{eq:F}\\
  \partial_i E^i &= \frac{1}{\epsilon_0}\rho,
  \label{eq:G}\\
  \partial_i B^i &= 0,\label{eq:M}
\end{align}
where $\varepsilon_{ijk}$ is the Levi-Civita symbol and $E_i$ and $B_i$ are the
electric field and magnetic flux density, respectively.
$\epsilon_0$ and $\mu_0$ are the permittivity and the magnetic permeability in
vacuum, respectively, which satisfy the relationship $\epsilon_0\mu_0=1/c^2$,
where $c$ is the speed of light.
$\rho$ and $J_i$ are the charge density and current density, respectively, which
satisfy the equation of continuity
\begin{align}
  \partial_t\rho + \partial_iJ^i=0.
  \label{eq:contE}
\end{align}

The Lagrangian density of the Maxwell's equations is defined as (see, e.g.,
\cite{Jackson, Low, Wald})
\begin{align}
  \mathcal{L}
  &
  \equiv \frac{\epsilon_0}{2}(E_iE^i-c^2B_iB^i)- \rho\phi + J^i A_i
  \nonumber\\
  &=
  \frac{\epsilon_0}{2}(\partial_i\phi+\partial_tA_i)
  (\partial^i\phi+\partial_tA^i)
  - \frac{1}{2\mu_0}(\partial^aA^b-\partial^bA^a)(\partial_aA_b)
  -\rho\phi + J^iA_i,
  \label{eq:Lagrangian}
\end{align}
where $\phi$ is the scalar potential and $A_i$ is the vector potential.
$\phi$ and $A_i$ satisfy below the relationship
\begin{align}
  E_i = -\partial_i\phi - \partial_tA_i,\quad
  B_i = \varepsilon_i{}^{mn}\partial_mA_n.
\end{align}
Then, we obtain the Hamiltonian density by the Legendre transformation (see,
e.g., \cite{Wald}) as,
\begin{align}
  \mathcal{H}
  &\equiv
  \pi\partial_t\phi+\Pi^i\partial_tA_i
  - \mathcal{L}\nonumber\\
  &=
  \phi(\rho  + \partial_i \Pi^i)
  + \frac{1}{2\epsilon_0}\Pi_i\Pi^i
  + \frac{1}{2\mu_0}(\partial^{a}A^{b} - \partial^{b}A^{a})(\partial_aA_b)
  -J^iA_i,
  \label{eq:Ham}
\end{align}
where $\pi\equiv (\delta\mathcal{L})/(\delta\partial_t\phi)$ and
$\Pi^i\equiv(\delta\mathcal{L})/(\delta\partial_tA_i)$ are the conjugate
momenta of $\phi$ and $A_i$, respectively.
They are written explicitly as
\begin{align}
  \pi &= 0,\\
  \Pi^i &= \epsilon_0(\partial^i\phi + \partial_tA^i).
\end{align}
Since $\pi=0$ identically, $\phi$ is a gauge variable.
Therefore, the variation of $\mathcal{H}$ with $\phi$ is a constraint equation,
i.e.,
\begin{align}
  0=-\frac{\delta\mathcal{H}}{\delta\phi}=-\rho - \partial_i\Pi^i,
\end{align}
Consequently, the canonical formulation of the Maxwell's equations is derived as
\begin{align}
  \mathcal{C}
  &\equiv
  - \rho - \partial_i\Pi^i,\label{eq:HamG}\\
  \partial_tA_i
  &\equiv \frac{\delta\mathcal{H}}{\delta \Pi^i}
  =-\partial_i\phi + \frac{1}{\epsilon_0}\Pi_i,
  \label{eq:HamA}\\
  \partial_t\Pi^i
  &\equiv -\frac{\delta\mathcal{H}}{\delta A_i}
  =\frac{1}{\mu_0}\partial_j\partial^jA^i
  - \frac{1}{\mu_0}\partial_j\partial^iA^j
  + J^i.
  \label{eq:HamPi}
\end{align}
The above set of equations is well-known the canonical formulation of the
Maxwell's equations \cite{PP}(There are other formulations, e.g.,\cite{AA}).
The details of the derivations are in \ref{sec:DerivationCanonicalEqs}.

Equation \eqref{eq:HamG} is consistent with Gauss's law \eqref{eq:G}, and
\eqref{eq:HamPi} is consistent with the Amp\`ere law with Maxwell's correction
\eqref{eq:AM}.
Using \eqref{eq:contE}, \eqref{eq:HamA}, and \eqref{eq:HamPi}, the evolution
equation of constraint $\mathcal{C}$ is calculated as
\begin{align}
  \partial_t\mathcal{C}
  &= -\partial_t\rho - \partial_t\partial_i\Pi^i
  \nonumber\\
  &= \partial_iJ^i
  - \partial_i\left(
  \frac{1}{\mu_0}\partial_j\partial^jA^i
  - \frac{1}{\mu_0}\partial^i\partial_jA^j
  + J^i
  \right)
  \nonumber\\
  &= 0,
  \label{eq:CP}
\end{align}
where the second equality is obtained using \eqref{eq:contE} and
\eqref{eq:HamPi}.
From \eqref{eq:CP}, \eqref{eq:HamG} is guaranteed regardless of time.
Hereafter, we call \eqref{eq:CP} the constraint propagation equation.
Note that \eqref{eq:CP} is not derived from the functional derivative using the
Hamiltonian density or the Lagrangian density.

\section{Discrete Variational Derivative Method
  \label{sec:DVDM}
}

Now we review the processes of deriving discretized equations using the DVDM,
and further details of the DVDM are given in \cite{FM}.
The discrete value of the variable $u$ is defined as $u{}^{(n)}_{(k)}$, where
the upper index $(n)$ and lower index $(k)$ denote the time component and space
component, respectively.
The forward and backward difference operators are defined as
$\widehat{\delta}^+_i u{}^{(n)}_{(k)}\equiv\bigl(u^{(n)}_{(k+1)}
- u^{(n)}_{(k)}\bigr)/\Delta x^i$ and $\widehat{\delta}^-_i u{}^{(n)}_{(k)}
\equiv\bigl(u^{(n)}_{(k)} - u^{(n)}_{(k-1)}\bigr)/\Delta x^i$, respectively,
and the central difference operator is defined as $\widehat{\delta}^\lra{1}_i
\equiv (\widehat{\delta}^+_i+\widehat{\delta}^-_i)/2$.
The second-order central difference operator is defined as
\begin{align*}
  \widehat{\delta}^{\langle2\rangle}_{ij}u^{(n)}_{(k)}\equiv
  \left\{
  \begin{array}{ll}
    (u^{(n)}_{(k+1)} - 2u^{(n)}_{(k)} + u^{(n)}_{(k-1)})/(\Delta x^i)^2,
    & (i=j)\\
    \widehat{\delta}^{\langle1\rangle}_i
    \widehat{\delta}^{\langle1\rangle}_j
    u^{(n)}_{(k)}. & (i\neq j)
  \end{array}
  \right.
\end{align*}
The symbol $\Delta x^i$ denotes $(\Delta x, \Delta y, \Delta z)$.
Now, we show the standard process for discretizing equations.
We adopt only the canonical formalism in this article, so that this process is
only shown in the case that the generator function is the Hamiltonian density.
The ordinary case is as follows:
(A) We set a Hamiltonian density $\mathcal{H}(q_i, p^i)$ with variables
$q_i$ and canonical conjugate momenta $p^i$.
(B) The dynamical equations are given as
\begin{align}
  \partial_t q_i&=\frac{\delta \mathcal{H}}{\delta p^i},\quad
  \partial_t p^i=-\frac{\delta \mathcal{H}}{\delta q_i},
\end{align}
and the discretized equations are constructed using a well-known method such as
the Crank-Nicolson scheme.
On the other hand, the process for the DVDM is as follows:
(A$'$) We set a discrete Hamiltonian density $\mathcal{H}^{(n)}_{(k)}\bigl(
q_i{}^{(n)}_{(k)}, p^i{}^{(n)}_{(k)}\bigr)$ with discrete variables
$q_i{}^{(n)}_{(k)}$ and discrete canonical conjugate momenta
$p^i{}^{(n)}_{(k)}$.
(B$'$) The discretized system obtained using the DVDM scheme is
\begin{align}
  \begin{array}{l}
    \displaystyle{\frac{q_i{}^{(n+1)}_{(k)}-q_{i}{}^{(n)}_{(k)}}{\Delta t}
      =\frac{\widehat{\delta} \mathcal{H}}{\widehat{\delta}(p^i{}^{(n+1)}_{(k)},
        p^i{}^{(n)}_{(k)})}},\\
    \displaystyle{\frac{p^i{}^{(n+1)}_{(k)}-p^{i}{}^{(n)}_{(k)}}{\Delta t}
      =-\frac{\widehat{\delta}\mathcal{H}}{\widehat{\delta}(q_i{}^{(n+1)}_{(k)},
        q_i{}^{(n)}_{(k)})}},
  \end{array}
\end{align}
where $\widehat{\delta} \mathcal{H}/\bigl(\widehat{\delta}(p^i{}^{(n+1)}_{(k)},
p^i{}^{(n)}_{(k)})\bigr)$ and $\widehat{\delta} \mathcal{H}/\bigl(
\widehat{\delta}(q_i{}^{(n+1)}_{(k)},q_i{}^{(n)}_{(k)})\bigr)$ are calculated as
\begin{align}
  \mathcal{H}^{(n+1)}_{(k)} - \mathcal{H}^{(n)}_{(k)}
  &=
  \frac{\widehat{\delta} \mathcal{H}}{\widehat{\delta}(p^i{}^{(n+1)}_{(k)},
    p^i{}^{(n)}_{(k)})}(p^{i}{}^{(n+1)}_{(k)}-p^{i}{}^{(n)}_{(k)})
  \nonumber\\
  &\quad
  + \frac{\widehat{\delta} \mathcal{H}}{\widehat{\delta}(q_i{}^{(n+1)}_{(k)},
    q_i{}^{(n)}_{(k)})}(q_{i}{}^{(n+1)}_{(k)}-q_{i}{}^{(n)}_{(k)}).
\end{align}
The discrete value $\mathcal{H}^{(n)}_{(k)}$ is not defined uniquely;
we can obtain different discretized equations depending on the definition.
Thus, we should select an appropriate definition to obtain good simulation
results.

\section{Discrete Maxwell's equations
  \label{sec:discreteMaxwellSystems}
}

In this section, we discretize the Maxwell's equations in three ways.
One way uses the iterative Crank-Nicolson scheme (ICNS), the others use the
DVDM.
We select the ICNS for comparing to the discretized equations using the DVDM.
This is because that we can see clearly and easily the differences between the
equations of ICNS and the ones of DVDM in the mathematical equations level.
There are some degrees of freedom when deriving discretized equations by the
DVDM,
and we derive two sets of discretized equations.
The first satisfies the constraint propagation equation at the discrete level,
which we call System I.
The other does not satisfy the equation, which we call System II.

\subsection{Iterative Crank-Nicolson Scheme}

The ICNS is one of the commonly used schemes to obtain discretized equations
from (partial) differential equations.
If we use this scheme, the evolution equations are discretized as the central
difference in time and space.
To obtain values at time step $n+1$ from those at time step $n$, it is used to
do two iterations \cite{Teukolsky00}.
Then, the discretized Maxwell's equations obtained using the ICNS are
\begin{align}
  \mathcal{C}^{(n)}_{(k)}
  &= -\rho^{(n)}_{(k)} - \widehat{\delta}^\lra{1}_i\Pi^i{}^{(n)}_{(k)},
  \label{eq:CNG}\\
  \frac{A_i{}^{(n+1)}_{(k)}-A_i{}^{(n)}_{(k)}}{\Delta t}
  &= - \frac{1}{2}\widehat{\delta}^\lra{1}_i\left(\phi^{(n+1)}_{(k)}
  +\phi^{(n)}_{(k)}\right)
  + \frac{1}{2\epsilon_0}\left(\Pi_i{}^{(n+1)}_{(k)}+\Pi_i{}^{(n)}_{(k)}
  \right),\label{eq:CNA}\\
  \frac{\Pi^i{}^{(n+1)}_{(k)} - \Pi^i{}^{(n)}_{(k)}}{\Delta t}
  &=
  \frac{1}{2\mu_0}\widehat{\delta}^{\langle2\rangle}_j{}^j
  \left(A^i{}^{(n+1)}_{(k)} + A^i{}^{(n)}_{(k)}\right)
  - \frac{1}{2\mu_0}\widehat{\delta}^{\langle2\rangle}_{j}{}^i
  \left(A^j{}^{(n+1)}_{(k)} + A^j{}^{(n)}_{(k)}\right)
  \nonumber\\
  &\quad
  + \frac{1}{2}\left(J^i{}^{(n+1)}_{(k)} + J^i{}^{(n)}_{(k)}\right),
  \label{eq:CNPi}
\end{align}
and the equation of continuity is discretized using the ICNS as
\begin{align}
  \frac{\rho^{(n+1)}_{(k)}-\rho^{(n)}_{(k)}}{\Delta t}
  &= - \frac{1}{2}\widehat{\delta}^\lra{1}_i\left(J^i{}^{(n+1)}_{(k)}
  + J^i{}^{(n)}_{(k)}\right).
  \label{eq:CNcontE}
\end{align}
Using these equations, we calculate the discretized evolution equation of
constraint $\mathcal{C}$ as
\begin{align}
  \frac{\mathcal{C}{}^{(n+1)}_{(k)}-\mathcal{C}{}^{(n)}_{(k)}}{\Delta t}
  &=
  -\frac{\rho{}^{(n+1)}_{(k)}-\rho{}^{(n)}_{(k)}}{\Delta t}
  - \widehat{\delta}^\lra{1}_i
  \frac{\Pi^i{}^{(n+1)}_{(k)}-\Pi^i{}^{(n)}_{(k)}}{\Delta t}
  \nonumber\\
  &=
  -\frac{1}{2\mu_0}\widehat{\delta}^{\langle1\rangle}_i
  \biggl\{\widehat{\delta}^{\langle2\rangle}_j{}^j
  (A^i{}^{(n+1)}_{(k)} + A^i{}^{(n)}_{(k)})
  \nonumber\\
  &\qquad
  - \widehat{\delta}^{\langle2\rangle}_j{}^i
  (A^j{}^{(n+1)}_{(k)} + A^j{}^{(n)}_{(k)})
  \biggr\}.
  \label{eq:CNCP}
\end{align}
The right-hand-side (r.h.s.) of this equation is not zero in general.
For instance, if we set $\partial_iA_j=0\,\,(i\neq j)$ as the initial
conditions, then the r.h.s. of \eqref{eq:CNCP} is zero.
On the other hand, if we set $\partial_iA_j\neq 0\,\,(i\neq j)$ as the initial
conditions, then the r.h.s of \eqref{eq:CNCP} is not zero.
We will confirm the results that the simulations using ICNS may not work
depending on the initial conditions by performing some simulations as reported
in Sec \ref{sec:Numericaltests}.

\subsection{System I}
To derive discretized equations by the DVDM, we have to define the discretized
Hamiltonian density appropriately.
In this article, we define the discretized Hamiltonian density
$\mathcal{H}^{(n)}_{(k)}$ as
\begin{align}
  \mathcal{H}^{(n)}_{(k)}
  &\equiv
  - \Pi^i{}^{(n)}_{(k)}(\widehat{\delta}^{\lra{1}}_i\phi^{(n)}_{(k)})
  + \frac{1}{2\epsilon_0}\Pi_i{}^{(n)}_{(k)}\Pi^i{}^{(n)}_{(k)}
  \nonumber\\
  &\quad
  + \frac{1}{2\mu_0}(\widehat{\delta}^{\lra{1}}{}^{i}A^{j}{}^{(n)}_{(k)}
  - \widehat{\delta}^{\lra{1}}{}^{j}A^{i}{}^{(n)}_{(k)})
  ({\delta}^{\lra{1}}_iA_j{}^{(n)}_{(k)})
  + \rho^{(n)}_{(k)}\phi^{(n)}_{(k)}
  - J_i{}^{(n)}_{(k)}A^i{}^{(n)}_{(k)}.
  \label{eq:DVHam}
\end{align}
Then the discretized Maxwell's equations are derived using the DVDM as
\begin{align}
  \mathcal{C}^{(n)}_{(k)}&=
  - \rho^{(n)}_{(k)}- \widehat{\delta}^{\lra{1}}_i \Pi^i{}^{(n)}_{(k)},
  \label{eq:DVG}
  \\
  \frac{A_i{}^{(n+1)}_{(k)}-A_i{}^{(n)}_{(k)}}{\Delta t}
  &=
  - (\widehat{\delta}^{\lra{1}}_i\phi^{(n+1)}_{(k)})
  + \frac{1}{2\epsilon_0}(\Pi_i{}^{(n+1)}_{(k)} + \Pi_i{}^{(n)}_{(k)}),
  \label{eq:DVA}\\
  \frac{\Pi^i{}^{(n+1)}_{(k)} - \Pi^i{}^{(n)}_{(k)}}{\Delta t}
  &=
  \frac{1}{2\mu_0}\widehat{\delta}^{\lra{1}}{}_j\widehat{\delta}^{\lra{1}}{}^j
  (A^i{}^{(n+1)}_{(k)}+A^i{}^{(n)}_{(k)})
  \nonumber\\
  &\quad
  - \frac{1}{2\mu_0}\widehat{\delta}^{\lra{1}}_j\widehat{\delta}^{\lra{1}}{}^i
  (A^j{}^{(n+1)}_{(k)}+A^j{}^{(n)}_{(k)})
  + J^i{}^{(n)}_{(k)}.
  \label{eq:DVPi}
\end{align}
Equation \eqref{eq:contE} is not derived from the functional derivative using
the Hamiltonian density in the continuous case.
Therefore, the discretized equation of continuity is also not derived from the
DVDM; we define it as
\begin{align}
  \frac{\rho^{(n+1)}_{(k)} - \rho^{(n)}_{(k)}}{\Delta t}
  + \widehat{\delta}^\lra{1}_iJ^i{}^{(n)}_{(k)}=0.
  \label{eq:DVcontE}
\end{align}
Now, we show that $(\mathcal{C}^{(n+1)}_{(k)}-\mathcal{C}^{(n)}_{(k)})/\Delta t$
is always zero independent of the initial conditions.
The discretized evolution of $\mathcal{C}^{(n)}_{(k)}$ is described by
\begin{align}
  \frac{\mathcal{C}^{(n+1)}_{(k)}-\mathcal{C}^{(n)}_{(k)}}{\Delta t}
  &=
  -\frac{\rho^{(n+1)}_{(k)} - \rho^{(n)}_{(k)}}{\Delta t}
  - \widehat{\delta}^\lra{1}_{i}
  \frac{\Pi^i{}^{(n+1)}_{(k)} - \Pi^i{}^{(n)}_{(k)}}{\Delta t}
  \nonumber\\
  &=
  \widehat{\delta}^{\langle1\rangle}_iJ^i{}^{(n)}_{(k)}
  - \widehat{\delta}^\lra{1}_{i}
  \biggl\{
  \frac{1}{2\mu_0}\widehat{\delta}^\lra{1}_j
  \widehat{\delta}^\lra{1}{}^j(A^i{}^{(n+1)}_{(k)} + A^i{}^{(n)}_{(k)})
  \nonumber\\
  &\qquad
  - \frac{1}{2\mu_0}\widehat{\delta}^\lra{1}_j
  \widehat{\delta}^\lra{1}{}^i(A^j{}^{(n+1)}_{(k)} + A^j{}^{(n)}_{(k)})
  + J^i{}^{(n)}_{(k)}
  \biggr\}\nonumber\\
  &=0,
  \label{eq:DVCP}
\end{align}
where the second equality is obtained using \eqref{eq:DVPi} and
\eqref{eq:DVcontE}.
Equation \eqref{eq:DVCP} indicates that $\mathcal{C}^{(n)}_{(k)}$ does not
change during the evolution.
Therefore, we claim that the equations discretized using System I is better than
that discretized using the ICNS.

\subsection{System II}
At the continuous level, the equation of continuity \eqref{eq:contE} is not
derived from the Hamiltonian density \eqref{eq:Ham}.
In the same way as at the discretized level, the discretized equation of
continuity is also not derived.
Thus, there are some degrees of freedom in deriving the discretized equations.
Thus, for instance, if we replace $J^i{}^{(n)}_{(k)}$ by $J^i{}^{(n+1)}_{(k)}$
in \eqref{eq:DVcontE}:
\begin{align}
  \frac{\rho^{(n+1)}_{(k)} - \rho^{(n)}_{(k)}}{\Delta t}
  + \widehat{\delta}^\lra{1}_iJ^i{}^{(n+1)}_{(k)}&=0.
  \label{eq:DVcontE2}
\end{align}
Using \eqref{eq:DVG}--\eqref{eq:DVPi} and \eqref{eq:DVcontE2}, the discretized
constraint propagation equation is
\begin{align}
  \frac{\mathcal{C}^{(n+1)}_{(k)}-\mathcal{C}^{(n)}_{(k)}}{\Delta t}
  &=-\frac{\rho^{(n+1)}_{(k)} - \rho^{(n)}_{(k)}}{\Delta t}
  - \widehat{\delta}^\lra{1}_iJ^i{}^{(n)}_{(k)}\nonumber\\
  &=-\widehat{\delta}^\lra{1}_i(J^i{}^{(n)}_{(k)}-J^i{}^{(n+1)}_{(k)}).
  \label{eq:DVCP2}
\end{align}
This equation is NOT equal to zero in general.
Therefore, we expect that the results of the simulations with System II will
become unstable if the divergence of $J_i$ changes during the evolution.

To modify System II, the use of a new Hamiltonian density is possible.
The result of the analysis of the discretized constraint propagation equation
in the modified System II is only the same as that for System I, as shown in
\ref{sec:modifyedSystemII}.

\section{Numerical Tests
  \label{sec:Numericaltests}
}

In this section, using the ICNS and the two types of the DVDM: System I and
System II, we perform some simulations in which we use two exact solutions of
the Maxwell's equations as the initial conditions.
In these simulations, we take the permittivity and magnetic permeability as
$\epsilon_0=\mu_0=1$ and the numerical parameters are set as follows:
\begin{itemize}
\item Simulation domain: $x,y,z\in[-0.5,0.5]$.
\item Grid: $x_i=y_i=z_i=-0.5+(i-(1/2))dx$, $i=1,\dots,100$,
  where $dx=1/100$.
\item Time step: $dt=0.1dx$.
\item Iteration: Second-order iterative calculations in ICNS, System I, and
  System II.
\item Gauge condition: Exact solution.
\item Boundary condition: Periodic boundary condition.
\end{itemize}
The constraint preserving character of System I would be independent of the
boundary condition, because that the discretized constraint propagation
equation \eqref{eq:DVCP} is calculated without the boundary condition.
In these simulations, since the characters of the scheme should be
distinguished the ones caused by other conditions, we adopt only the boundary
condition periodically.

The accuracies of the three schemes are all of second order in times and
space, because that the difference operators $\delta^{\lra{1}}_i$ and
$\delta^{\lra{2}}_{ij}$ are linear and second order in space.
In addition, we confirm it in numerically with an initial data in
\ref{sec:Convergence}.

\subsection{Case 1}

\begin{figure}[htbp]
  \centering
  \includegraphics[keepaspectratio=true,width=0.8\hsize]{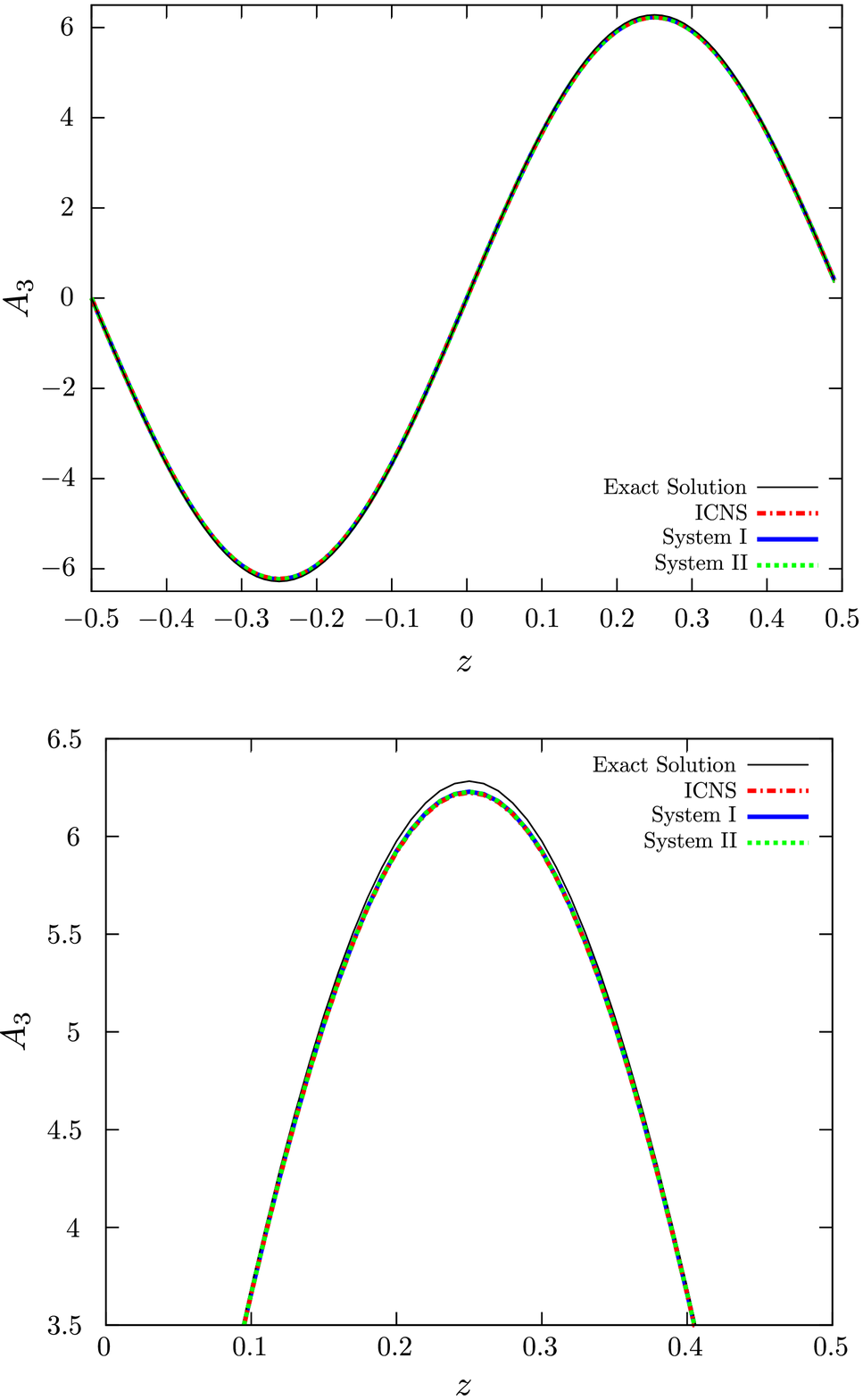}
  \caption{\label{fig:WavesA3CaseI}
    These lines express the solutions of $A_3$ at $t=5$.
    The vertical axis is $A_3$ and the horizontal axis is $z$.
    The top panel is drawn in $-0.5\leq z\leq 0.5$ and
    $-6.5\leq A_3\leq 6.5$, and the bottom panel is the one in
    $0\leq z\leq 0.5$ and $3.5\leq A_3\leq6.5$.
    The thin line is drawn the exact solution, the dot-dashed line is using
    ICNS, the thick line is using System I, and the dotted line is using
    System II.
    The three lines except the thin line are overlapping.
  }
\end{figure}

First, we use the following exact solution of the Maxwell's equations as the
initial condition:
\begin{align}
  A_i&=\frac{2\pi}{\omega^2}\begin{pmatrix}
    \cos(\omega t+2\pi x)\\
    \sin(\omega t+2\pi y)\\
    \omega^2\sin(2\pi z)
  \end{pmatrix},\label{eq:IniA1}\\
  \Pi^i
  &=\frac{2\pi}{\omega}\begin{pmatrix}
    -\sin(\omega t+2\pi x)\\
    \cos(\omega t+2\pi y)\\
    -\omega t\sin(2\pi z)
  \end{pmatrix},\\
  \phi &= t\cos(2\pi z),\label{eq:IniPhi1}\\
  \rho &= \frac{4\pi^2}{\omega}\{
  \cos(\omega t+2\pi x)
  + \sin(\omega t+2\pi y)
  + \omega t\cos(2\pi z)\},\\
  J_i &=
  -2\pi\begin{pmatrix}
  \cos(\omega t+2\pi x)\\
  \sin(\omega t+2\pi y)\\
  \sin(2\pi z)
  \end{pmatrix},\label{eq:IniJ1}
\end{align}
where $\omega=1$ and $t=1$.
With this initial condition, the discretized constraint propagation equation of
the ICNS \eqref{eq:CNCP} become zero because the vector potential $A_i$
satisfies the condition $\partial_iA_j=0\quad(i\neq j)$.
Moreover, the divergence of $J_i$ is
\begin{align*}
  \partial_iJ^i=4\pi^2(\sin(\omega t+2\pi x)-\cos(\omega t+2\pi y)
  - \cos (2\pi z)),
\end{align*}
and the discretized constraint propagation equation for System II
\eqref{eq:DVCP2} does not always become zero in the region
$[-0.5,0.5]\times[-0.5,0.5]\times[-0.5,0.5]$.
\begin{figure}[htbp]
  \centering
  \includegraphics[keepaspectratio=true,width=\hsize]{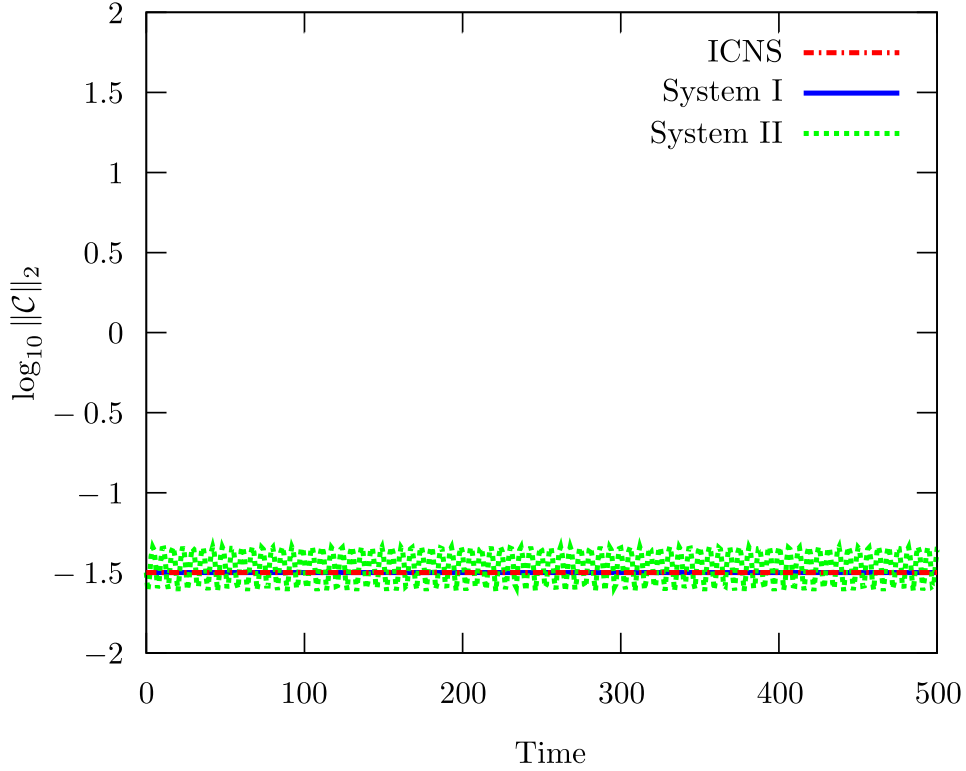}
  \caption{\label{fig:Case1}
    L2 norm of constraint $\mathcal{C}$ for the ICNS and the two types of DVDM
    (System I and System II).
    The vertical axis is the logarithm of $\mathcal{C}$ and the horizontal axis
    is time.
    The dot-dashed line represents the ICNS, the solid line represents the
    System I, and the dotted line represents the System II.
    The lines for the ICNS and System I overlap throughout the evolution.
  }
\end{figure}
\begin{figure}[htbp]
  \centering
  \includegraphics[keepaspectratio=true,width=\hsize]{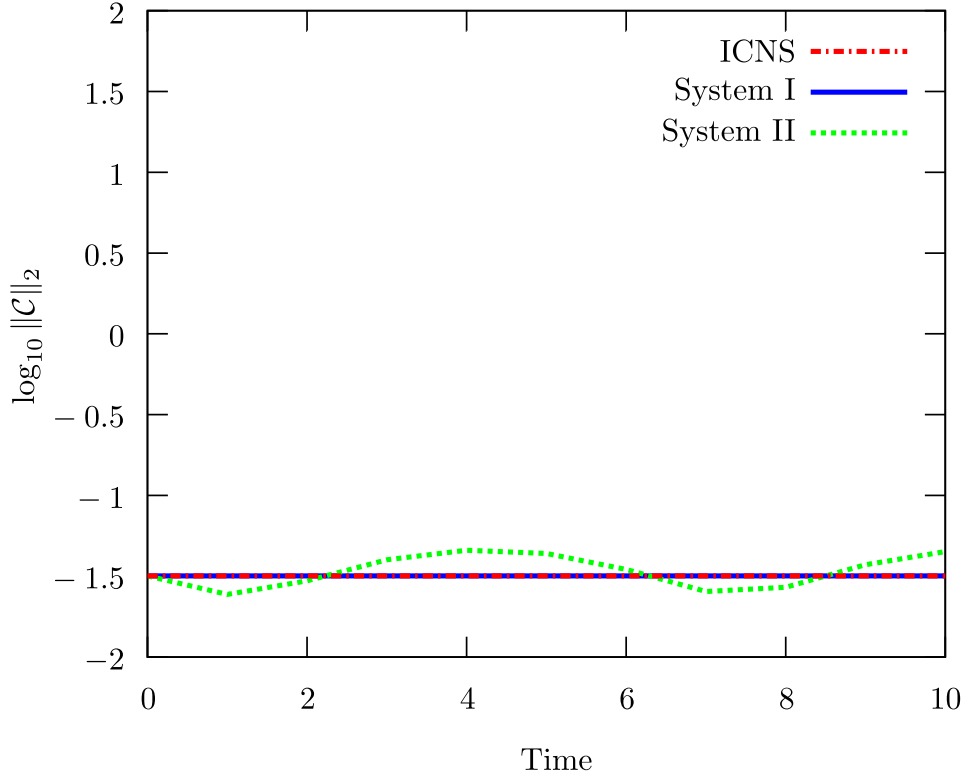}
  \caption{\label{fig:Case1-2}
    L2 norm of $\mathcal{C}$ drawn up to $t=10$.
    All numerical conditions are the same as those for Figure \ref{fig:Case1}.
  }
\end{figure}
Figure \ref{fig:Case1} shows the L2 norm of constraint $\mathcal{C}$ for the
ICNS \eqref{eq:CNG}--\eqref{eq:CNcontE}, System I
\eqref{eq:DVG}--\eqref{eq:DVcontE}, and System II
\eqref{eq:DVG}--\eqref{eq:DVPi}, \eqref{eq:DVcontE2}.
Figure \ref{fig:Case1-2} shows details of the interval up to $t=10$ in
Figure \ref{fig:Case1}.
We see that the two lines for the ICNS and System I are unchanged during the
evolution, whereas that for System II changes with time.
These results are consistent with the previous analysis of the discretized
constraint propagation equations.
Figure \ref{fig:WavesA3CaseI} shows the solutions of $A_3$ of $t=5$,
$-0.5\leq z\leq 0.5$ in three schemes.
The bottom panel of the figures shows the details of the range
$3.5\leq A_3\leq 6.5$ in the top panel.
In the figures, we see three lines using ICNS, System I, and System II are
overlapping.
\begin{figure}[htbp]
  \centering
  \includegraphics[keepaspectratio=true,width=0.8\hsize]{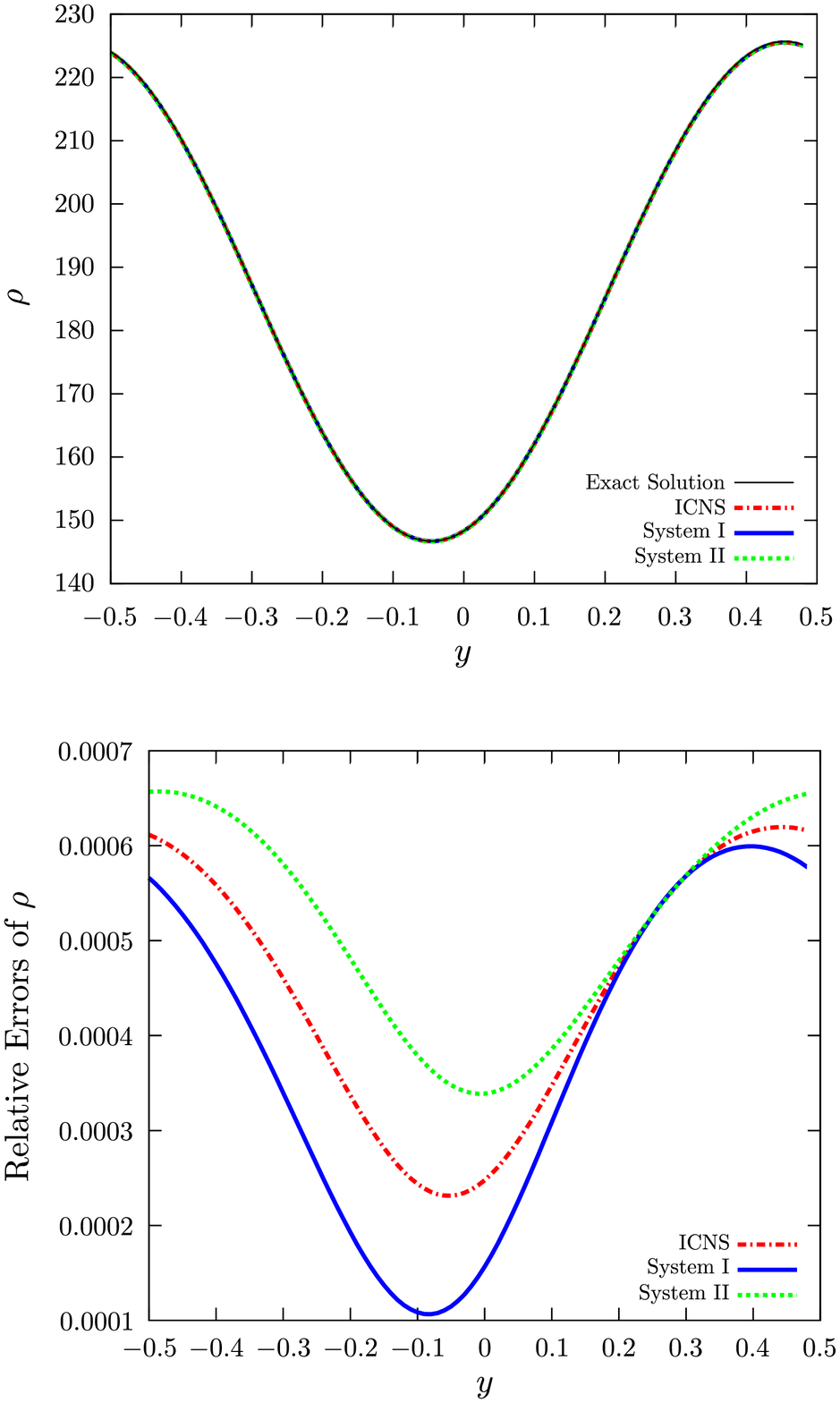}
  \caption{\label{fig:WavesRhoCaseI}
    These lines express the solutions of $\rho$ at $t=5$, $x=-0.5$,
    $z=0$.
    The top panel is drawn in $-0.5\leq y\leq 0.5$ and
    $140\leq \rho\leq 230$.
    The thin line is drawn the exact solution, the dot-dashed line is using
    ICNS, the thick line is using System I, and the dotted line is using
    System II.
    All of the lines are almost overlapping.
    The bottom panel is drawn the relative errors of the three schemes
    against the exact solution.
    The dot-dashed line is drawn the relative errors of ICNS, the solid line
    is the ones of System I, and the dotted line is the ones of System II.
    }
\end{figure}
The top panel of Figure \ref{fig:WavesRhoCaseI} shows the solutions of $\rho$
of $t=5$, $x=-0.5$, $z=0$, $-0.5\leq y\leq 0.5$ in three schemes.
We see all the lines are almost overlapping.
To see the differences of the schemes more details, we show the relative
errors against the exact solution of the three schemes in the bottom panel of
Figure \ref{fig:WavesRhoCaseI}.
The differences between each value of the schemes hardly exist.
For other conditions such as $t=10$ and other variables such as $\Pi^i$, the
results are similar.

\subsection{Case 2}

\begin{figure}[htbp]
  \centering
  \includegraphics[keepaspectratio=true,width=0.8\hsize]{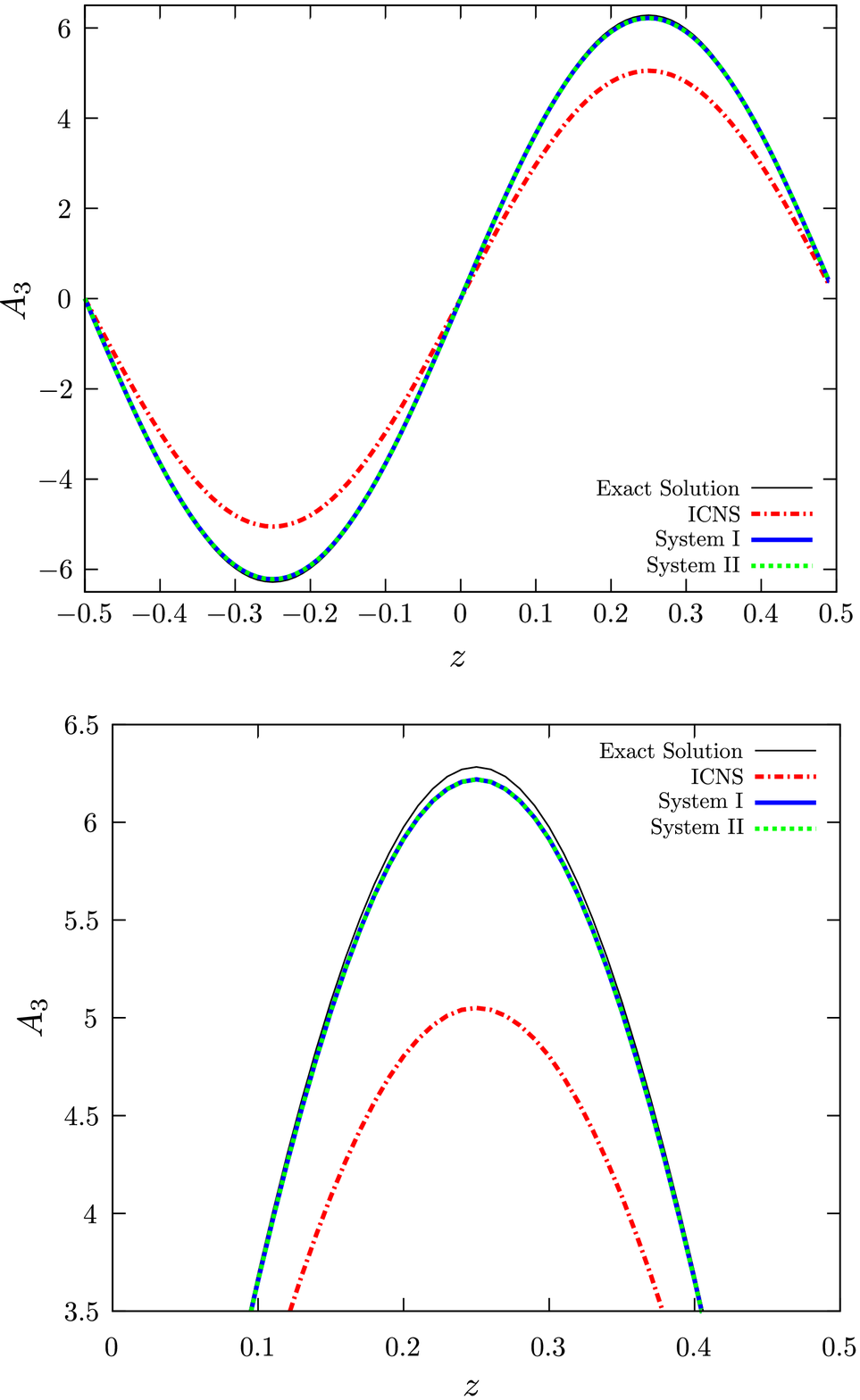}
  \caption{\label{fig:WavesA3CaseII}
    These lines express the solutions of $A_3$ at $x=y=-0.5$ in the same
    conditions of Figure \ref{fig:WavesA3CaseI} except the initial condition.
    The two lines using System I and System II are overlapping.
  }
\end{figure}

\begin{figure}[htbp]
  \centering
  \includegraphics[keepaspectratio=true,width=0.8\hsize]{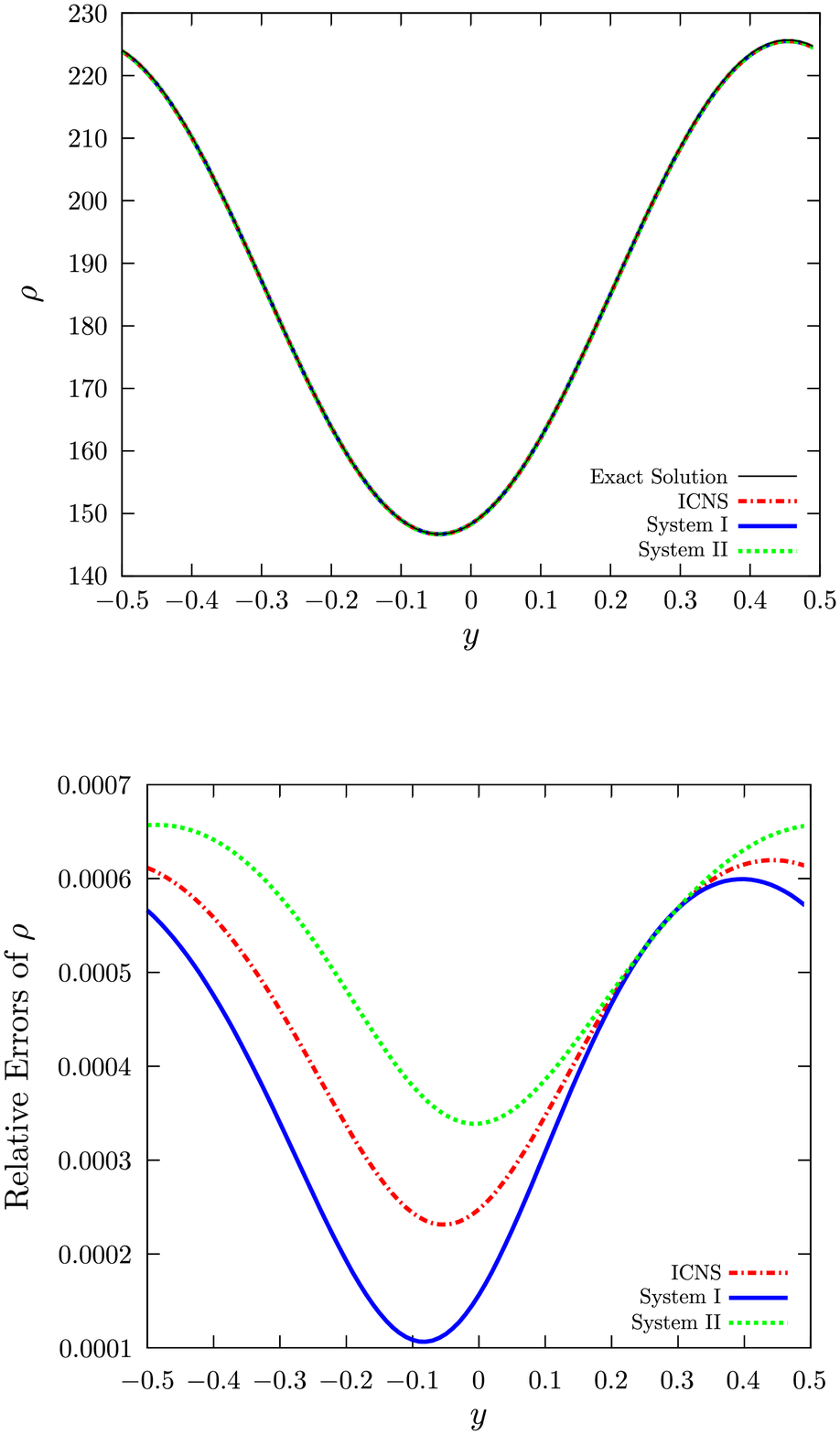}
  \caption{\label{fig:WavesRhoCaseII}
    These lines express the solutions of $\rho$ in the same conditions of
    Figure \ref{fig:WavesA3CaseII} except the initial condition.
    The all lines of the top panel are almost overlapping.
  }
\end{figure}

Next, we use the following exact solution as the initial condition:
\begin{align}
  A_i&=\frac{2\pi}{\omega^2}\begin{pmatrix}
    \cos(\omega t+2\pi x)\\
    \sin(\omega t+2\pi y)\\
    \omega^2\sin(2\pi (x+y+z))
  \end{pmatrix},\label{eq:IniA2}\\
  \Pi^i
  &=\frac{2\pi}{\omega}\begin{pmatrix}
    -\sin(\omega t+2\pi x)\\
    \cos(\omega t+2\pi y)\\
    -\omega t\sin(2\pi z)
  \end{pmatrix},\\
  \phi &= t\cos(2\pi z),\\
  \rho &= \frac{4\pi^2}{\omega}\{
  \cos(\omega t+2\pi x)
  + \sin(\omega t+2\pi y)
  + \omega t\cos(2\pi z)\},\\
  J_i &=
  -2\pi\begin{pmatrix}
  \cos(\omega t+2\pi x)+4\pi^2\sin(2\pi(x+y+z))\\
  \sin(\omega t+2\pi y)+4\pi^2\sin(2\pi(x+y+z))\\
  \sin(2\pi z)-8\pi^2\sin(2\pi(x+y+z))
  \end{pmatrix},\label{eq:IniJ2}
\end{align}
where $\omega=1$ and $t=1$.
The discretized constraint propagation equation with the ICNS \eqref{eq:CNCP}
and that with System II \eqref{eq:DVCP2} do not equal zero for this condition.
\begin{figure}[htbp]
  \centering
  \includegraphics[keepaspectratio=true,width=\hsize]{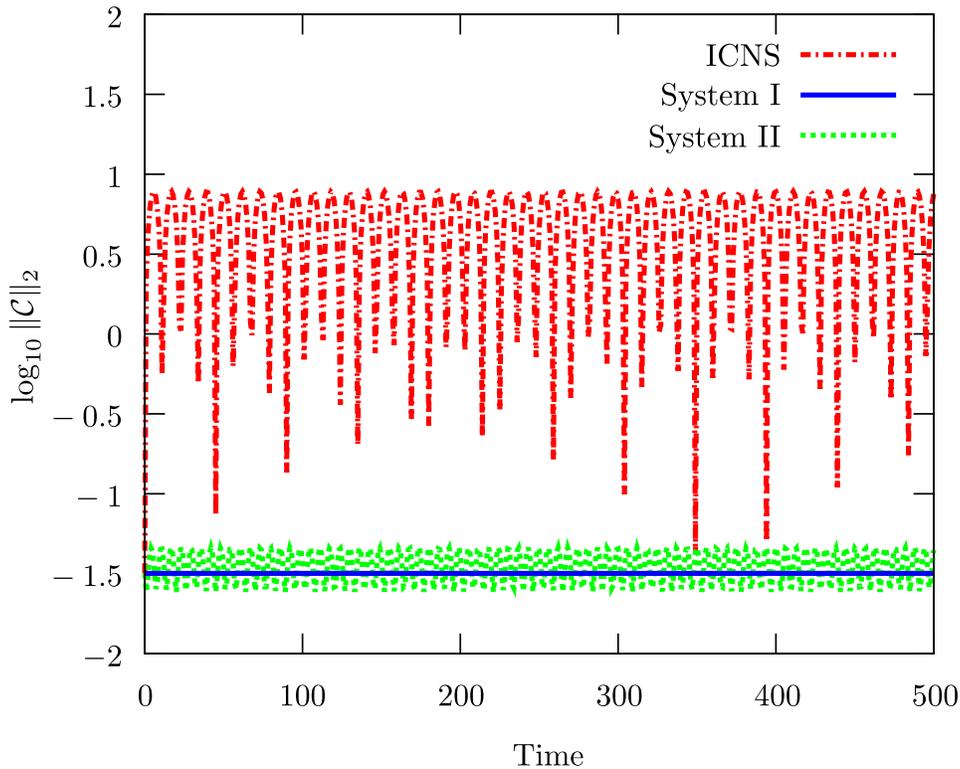}
  \caption{\label{fig:Case2}
    L2 norm of $\mathcal{C}$ for the same conditions as Figure \ref{fig:Case1}
    but with the initial condition given by \eqref{eq:IniA2}--\eqref{eq:IniJ2}.
    The lines representing the ICNS and System II oscillate and are unstable.
    On the other hand, the line for System I is unchanged from the initial
    state and is stable.
  }
\end{figure}
\begin{figure}[htbp]
  \centering
  \includegraphics[keepaspectratio=true,width=\hsize]{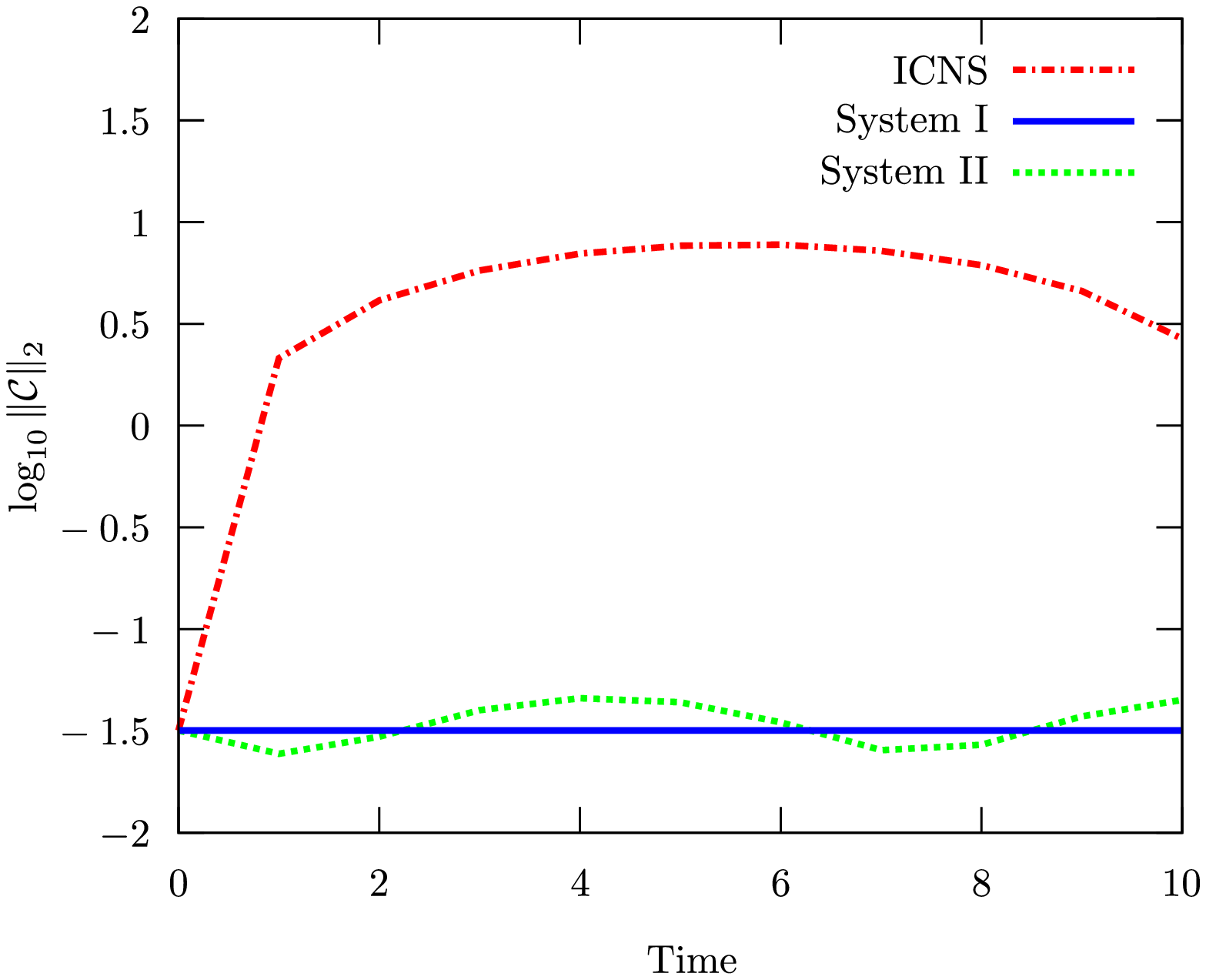}
  \caption{\label{fig:Case2-2}
    L2 norm of $\mathcal{C}$ is drawn up to $t=10$.
    All numerical conditions are the same as those for Figure \ref{fig:Case2}.
  }
\end{figure}
Figure \ref{fig:Case2} is drawn under the same conditions as Figure
\ref{fig:Case1} except for the initial condition. Figure \ref{fig:Case2-2}
shows details of the interval up to $t=10$ in Figure \ref{fig:Case2}.
We see that the constraint violations of System I do not change during the
evolution, meaning that the simulation is stable.
On the other hand, those of the ICNS and System II are unstable.
These results are consistent with the conclusions of the analysis at the
discretized level in Sec. \ref{sec:discreteMaxwellSystems}.
Figure \ref{fig:WavesA3CaseII} shows the solutions of $A_3$ at $x=y=-0.5$ and
Figure \ref{fig:WavesRhoCaseII} shows the solutions of $\rho$ in the same
conditions of Figure \ref{fig:WavesA3CaseI} and Figure
\ref{fig:WavesRhoCaseI}, respectively, except the initial condition.
Comparing the differences between the lines of the three schemes and the exact
solutions in Figure \ref{fig:WavesA3CaseII}, we see the difference of ICNS is
the largest.
This is consistent with the Figure \ref{fig:Case2} and Figure
\ref{fig:Case2-2}.
For the variables $A_1, A_2$ and $\Pi^i$, the results are similar with the
case of $A_3$.
In Figure \ref{fig:WavesRhoCaseII}, we see that the differences of the schemes
are hardly.

\section{Summary
  \label{sec:summary}
}

To perform accurate simulations, we apply the discrete variational derivative
method (DVDM) to the Maxwell's equations to derive a new set of discretized
Maxwell's equations.
In this process, we proposed a discretized Hamiltonian density and a discretized
equation of continuity, showed that the constraint is unchanged at the
discretized level, and confirmed this in some simulations.
Comparing the numerical solutions and the exact solutions, we assert that
the conservation of the constraint must be one of the important factors
to get the precise numerical solutions.
The definitions of the discretized Hamiltonian density and the discretized
equation of continuity are not unique and are restricted to satisfy the
constraint.
A way of discretizing the equation of continuity was not obtained from the DVDM
but from the analysis of the discretized constraint propagation equation.
Therefore, we claim that we have found a way to appropriately define the
discretized equation of continuity.
We conclude that the analysis of the discretized constraint propagation
equations is the key to perform accurate simulations in the constrained
dynamical system.

In this article, we studied the Maxwell's equations.
There are some constrained dynamical systems except for the Maxwell's equations.
For instance, the Einstein's field equations in general relativity is the one of
such systems.
Therefore we will construct appropriate discretized Einstein's field equations
by the DVDM in near the future.

\section*{Acknowledgements}
G. Y. was partially supported by a Waseda University Grant for Special Research
Projects (number 2015B-190).

\appendix

\section{Derivation of the canonical formulation of Maxwell's equations
  \label{sec:DerivationCanonicalEqs}
}
The Hamiltonian density is as \eqref{eq:Ham}:
\begin{align*}
  \mathcal{H}
  =
  \phi(\rho+\partial_i \Pi^i)
  +\frac{1}{2\epsilon_0}\Pi_i\Pi^i
  +\frac{1}{2\mu_0}(\partial^{a}A^{b} - \partial^{b}A^{a})(\partial_aA_b)
  -J^iA_i,
\end{align*}
then $A_i$, $\Pi^i$, and $\phi$ are independent variables.
Since the $\delta\mathcal{L}/\delta (\partial_t\phi)$ is identically zero,
$\phi$ is the gauge variable and the variation of $\phi$ is a constraint
equation \eqref{eq:HamG}:
\begin{align*}
  \mathcal{C}
  \equiv -\frac{\delta\mathcal{H}}{\delta \phi}
  = -\rho - \partial_i\Pi^i.
\end{align*}
The variations of $\Pi^i$ and $A_i$ are the evolution equations of $A_i$ and
$\Pi^i$, respectively, such as
\begin{align*}
  \partial_tA_i
  &\equiv \frac{\delta\mathcal{H}}{\delta \Pi^i}
  = \phi\partial_i + \frac{1}{\epsilon_0}\Pi_i
  = -\partial_i\phi + \frac{1}{\epsilon_0}\Pi_i
  + (\text{boundary terms}),\\
  \partial_t\Pi^i
  &\equiv -\frac{\delta\mathcal{H}}{\delta A_i}
  = -\frac{1}{2\mu_0}(\partial_aA_b)(\delta^{bi}\partial^a
  - \delta^{ai}\partial^b)
  - \frac{1}{2\mu_0}(\partial^aA^b + \partial^bA^a)\delta^i{}_b\partial_a
  + J^i\\
  &=
  \frac{1}{\mu_0}\partial_j\partial^jA^i
  - \frac{1}{\mu_0}\partial_j\partial^iA^j
  + J^i
  + (\text{boundary terms}),
\end{align*}
the boundary terms in the above equations can be vanished in a suitable
boundary condition, then we can get the equation \eqref{eq:HamA} and
\eqref{eq:HamPi}, respectively:
\begin{align*}
  \partial_tA_i
  &= -\partial_i\phi + \frac{1}{\epsilon_0}\Pi_i,\\
  \partial_t\Pi^i
  &=\frac{1}{\mu_0}\partial_j\partial^jA^i
  - \frac{1}{\mu_0}\partial_j\partial^iA^j
  + J^i.
\end{align*}

\section{Modified equations of System II
  \label{sec:modifyedSystemII}
}

To make the discretized constraint propagation equations of System II equal to
zero, we replace $J_i{}^{(n)}_{(k)}$ with $J_i{}^{(n+1)}_{(k)}$ in the
Hamiltonian density \eqref{eq:DVHam}; thus, we redefine the discretized
Hamiltonian density as
\begin{align}
  \bar{\mathcal{H}}^{(n)}_{(k)}
  &\equiv
  - \Pi^i{}^{(n)}_{(k)}(\widehat{\delta}^{\lra{1}}_i\phi^{(n)}_{(k)})
  + \frac{1}{2\epsilon_0}\Pi_i{}^{(n)}_{(k)}\Pi^i{}^{(n)}_{(k)}
  \nonumber\\
  &\quad
  + \frac{1}{\mu_0}(\widehat{\delta}^{\lra{1}}{}^{i}A^{j}{}^{(n)}_{(k)}
  - \widehat{\delta}^{\lra{1}}{}^{j}A^{i}{}^{(n)}_{(k)})
  (\widehat{\delta}^{\lra{1}}{}_{i}A_{j}{}^{(n)}_{(k)})
  + \rho^{(n)}_{(k)}\phi^{(n)}_{(k)}
  - J_i{}^{(n+1)}_{(k)}A^i{}^{(n)}_{(k)},
  \label{eq:DVHam2}
\end{align}
then the discretized Maxwell's equations are derived by the DVDM as
\eqref{eq:DVG} \eqref{eq:DVA} and
\begin{align}
  \frac{\Pi_i{}^{(n+1)}_{(k)} - \Pi_i{}^{(n)}_{(k)}}{\Delta t}
  &=
  J_i{}^{(n+1)}_{(k)}
  + \frac{1}{2\mu_0}\widehat{\delta}^{\lra{1}}{}_a\widehat{\delta}^{\lra{1}}{}^a
  (A_i{}^{(n+1)}_{(k)}+A_i{}^{(n)}_{(k)})
  \nonumber\\
  &\quad
  - \frac{1}{2\mu_0}\widehat{\delta}^{\lra{1}}_a\widehat{\delta}^{\lra{1}}_i
  (A^a{}^{(n+1)}_{(k)}+A^a{}^{(n)}_{(k)}).
  \label{eq:DVPi3}
\end{align}
We refer to \eqref{eq:DVG}, \eqref{eq:DVA}, \eqref{eq:DVPi3}, and
\eqref{eq:DVcontE2} as System III.
The discretized constraint propagation equation of System III is equal to zero.
The results of numerical simulations using System III were expected to be
the same as the those System I.
We performed some simulations using System III for Case 1 and Case 2 in Sec.
\ref{sec:Numericaltests} as the initial conditions and confirmed that the
constraint violations of System III are consistent with those of System I.

However, the Hamiltonian density  \eqref{eq:DVHam2} is unnatural because the
$(n+1)$th time component is included.
Therefore, we should define the discretized Hamiltonian density as
\eqref{eq:DVHam} and the discretized equation of continuity as
\eqref{eq:DVcontE}.

\section{Convergence Test of System I and System II in Case II
  \label{sec:Convergence}}
We show that the both of the convergences in the System I and System II are
second order of $\Delta x$ and $\Delta t$.
\begin{figure}[htbp]
  \centering
  \includegraphics[keepaspectratio=true,width=\hsize]{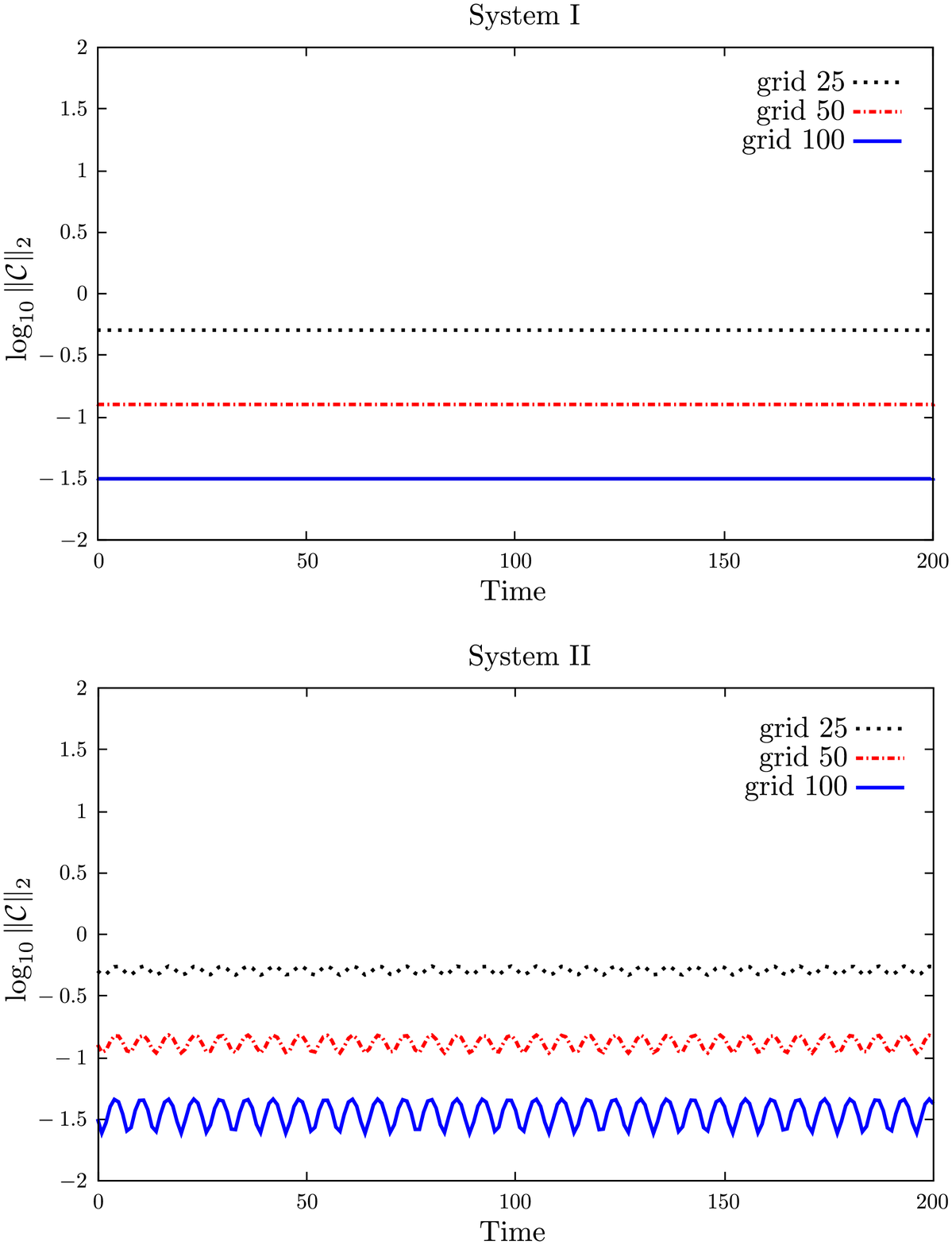}
  \caption{\label{fig:ConvergenceDVDMs}
    Upper panel is drawn using System I, lower panel is using System II.
    The vertical axis is the logarithm of $\mathcal{C}$ and the horizontal
    axis is time.
    The dotted line is drawn in the grid as $\Delta x=\Delta y=\Delta z
    =1/25$, the dot-dashed line drawn in $\Delta x=\Delta y=\Delta z=1/50$,
    and the solid line drawn in $\Delta x=\Delta y=\Delta z=1/100$.
    $\log_{10}\|\mathcal{C}\|_2$ of the dotted line is around $-0.3$, the
    one of the dot-dashed line is around $-0.9$, and the one of the dotted
    line is around $-1.5$.
  }
\end{figure}
The top panel and the bottom panel in Figure \ref{fig:ConvergenceDVDMs} are
drawn using System I and System II, respectively, in the initial condition as
Case II.
We can see the differences between $\log_{10}\|\mathcal{C}\|_2$ of the grids
are $0.6\approx \log_{10}4$, these results indicate that both of the
convergence properties of the systems are second order in time and space.


\end{document}